\documentclass[
preprint,
showpacs
]{revtex4-1}
\usepackage{amsthm}
\theoremstyle{plain}
\newtheorem{theorem}{Theorem}%[section]
\newtheorem*{theorem*}{Theorem}

\newtheorem*{definition*}{Definition}

\newtheorem*{lemma*}{Lemma}

\newcommand{\E}{\mathbb{E}}

\usepackage{braket}
\usepackage{delarray}
\usepackage{here}

\usepackage{amsmath}
\usepackage{txfonts}

\usepackage[dvipdfmx]{graphicx}
\usepackage{bm}
\usepackage{color}
\usepackage{ulem}

\usepackage{physics}

\newcommand{\ba}{\begin{array}}
\newcommand{\ea}{\end{array}}
\newcommand{\bmat}{\left(\begin{array}}
\newcommand{\emat}{\end{array}\right)}
\newcommand{\no}{\nonumber}

\newcommand{\be}{\begin{eqnarray}}
\newcommand{\ee}{\end{eqnarray}}

\begin{document}
\title{Toward Mean-Field Bound For Critical Temperature on Nishimori Line}
\author{Manaka Okuyama$^1$}
\author{Masayuki Ohzeki$^{1,2,3}$}
\affiliation{$^1$Graduate School of Information Sciences, Tohoku University, Sendai 980-8579, Japan}
\affiliation{$^2$Department of Physics, Tokyo Institute of Technology, Tokyo 152-8551, Japan}
\affiliation{$^3$Sigma-i Co., Ltd., Tokyo 108-0075, Japan} %\\

\begin{abstract}  
The critical inverse temperature of the mean-field approximation establishes a lower bound of the true critical inverse temperature in a broad class of ferromagnetic spin models. This is referred to as the mean-field bound for the critical temperature.
In this study, we explored the possibility of a corresponding mean-field bound for the critical temperature in Ising spin glass models with Gaussian randomness on the Nishimori line.
On this line, the critical inverse temperature of the mean-field approximation is given by $\beta_{MF}^{NL}=\sqrt{1/z}$, where $z$ is the coordination number.
Using the Griffiths inequality on the Nishimori line, we proved that there is zero spontaneous magnetization in the high-temperature region $\beta < \beta_{MF}^{NL}/2$. 
In other words, the true critical inverse temperature $\beta_c^{NL}$ on the Nishimori line is always bounded by $\beta_c^{NL} \ge \beta_{MF}^{NL}/2$. 
Unfortunately, we have not succeeded in obtaining the corresponding mean-field bound $\beta_c^{NL} \ge \beta_{MF}^{NL}$ on the Nishimori line.

\end{abstract}
\date{\today}
\maketitle

%%%%%%%%%%%%%%%%%%%%%%%%%%%%%%%%%%%%%%%%%%%%%%%%%%%%%%%%%%%%%%%%%%%%%%%%%%%
%%%%%%%%%%%%%%%%%%%%%%%%%%%%%%%%%%%%%%%%%%%%%%%%%%%%%%%%%%%%%%%%%%%%%%%%%%%
\section{Introduction}
The free energy of the mean-field approximation is always an upper bound of the true free energy.
This universal constraint on the free energy results from the Gibbs--Bogoliubov inequality~\cite{Kvasnikov,Bogoliubov}, which is based on the convexity of the pressure function $\log{Z}$ for the inverse temperature~\cite{Griffiths}.

For ferromagnetic models, additional constraints on the critical inverse temperature can be established.
It has been rigorously proved that the critical inverse temperature of the mean-field approximation sets a lower bound for the true critical inverse temperature in a broad class of ferromagnetic spin models. This is referred to as the mean-field bound for the critical temperature~\cite{Griffiths2,DLP,Simon,Sokal,Fisher}.
It is also known that the mean-field bound for magnetization holds~\cite{Thompson,Pearce,Slawny,Newman,TH}.
Unfortunately, the proofs of these bounds strongly rely on ferromagnetic interactions, making it seemingly impossible to extend them to spin glass models.

Concerning Ising spin glass models with Gaussian randomness on the Nishimori line~\cite{Nishimori,Nishimori2}, it was recently proved that the counterpart of the Gibbs--Bogoliubov inequality holds on the Nishimori line by using the convexity of the quenched pressure function $\mathbb{E}[\log{Z}]$ with respect to the parameters along the Nishimori line~\cite{OO}.
The Gibbs--Bogoliubov inequality on the Nishimori line shows that the quenched free energy of the mean-field approximation on the Nishimori line establishes an upper bound for the true quenched free energy on the Nishimori line~\cite{OO}.
This constraint on the quenched free energy of the mean-field approximation on the Nishimori line is notably similar to that of the conventional mean-field approximation resulting from the conventional Gibbs--Bogoliubov inequality.
Furthermore, Ising spin glass models on the Nishimori line behave similarly to ferromagnetic models, and various exact and rigorous results have been obtained~\cite{NS,MNC,Kitatani,ACCM,KM,BM,BM2,ACCM2,Ohzeki,ON2,Ohzeki2,KM,GS,Nishimori10,IS,IS2,Nishimori3,GHDB,ON,Ozeki,Ozeki2,Iba,GRL,MON,OO2,CPM}.
Then, from the mean-field bound for the critical temperature in ferromagnetic models, we have a natural question: how does the critical inverse temperature $\beta_{MF}^{NL}$ of the mean-field approximation on the Nishimori line relate to the true critical inverse temperature $\beta_c^{NL}$ on the Nishimori line? 
In particular, does the corresponding mean-field bound $\beta_c^{NL} \ge \beta_{MF}^{NL}$ hold on the Nishimori line?

In this study, we partially answered this question.
Using the Griffiths inequality on the Nishimori line~\cite{MNC}, we proved that there is always zero spontaneous magnetization in the high-temperature region $\beta < \beta_{MF}^{NL}/2$. 
This immediately implies that the true critical inverse temperature on the Nishimori line is always bounded by $\beta_C^{NL} \ge \beta_{MF}^{NL}/2$.
Unfortunately, it remains an open problem whether the corresponding mean-field bound on the Nishimori line, $\beta_c^{NL} \ge \beta_{MF}^{NL}$, holds.

The remainder of this paper is organized as follows.
In Sec. II, we define the model and present the main result (Theorem 1).
Section III presents the proof of Theorem 1. 
Finally, a discussion is presented in Sec. IV.

%%%%%%%%%%%%%%%%%%%%%%%%%%%%%%%%%%%%%%%%%%%%%%%%%%%%%%%%%%%%%%%%%
\section{Definitions and Results}
Let $\Lambda$ be any lattice invariant under translations with coordination number $z$ (for example, $\Lambda=Z^d$ with $z=2d$).
Free or fixed boundary conditions on the Nishimori line are imposed~\cite{MNC}.
To each site $i\in\Lambda$, we associate an Ising spin $\sigma_i=\pm1$.
The Hamiltonian of the Ising spin glass model with Gaussian randomness on the Nishimori line is defined as
\be
H&=&-\sum_{\langle i j\rangle}J_{ij} \sigma_i \sigma_j - \sum_{i} h_i \sigma_i,
\ee
where $\langle i j\rangle$ runs over interacting spin pairs on $\Lambda$, and $J_{ij}$ and $h_{i}$ are independent Gaussian random variables distributed as follows:
\be
J_{ij} \overset{\text{iid}}{\sim} \mathcal{N}\left(\beta, 1 \right) , \label{NL-condition}
\\
 h_{i} \overset{\text{iid}}{\sim} \mathcal{N}\left(\beta h, h \right) ,\label{NL-condition2}
\ee
where $h\ge0$, and $\beta$ is the inverse temperature.
The choice of the Gaussian distributions (\ref{NL-condition}) and (\ref{NL-condition2}) satisfies the condition of the Nishimori line~\cite{Nishimori2}.
We note that fixed boundary condition on the Nishimori line~\cite{MNC} is represented by taking the limit $h\to\infty$ for all sites outside $\Lambda$, without violating the condition of the Nishimori line.

The partition function is given by
\be
Z_{} &=&  \Tr \qty(e^{-\beta H}),
\ee
where $\Tr(\cdots)$ denotes the summation over all spin variables.
The quenched free energy on the Nishimori line in the thermodynamic limit is defined as
\be
-\beta f_{NL}&=&\lim_{|\Lambda|\to\infty}\frac{1}{|\Lambda|}\mathbb{E}\left[\log Z_{} \right],
\ee
where $|\Lambda|$ is the number of the sites, and $\mathbb{E}[\cdots]$ denotes the expectation with respect to all random variables.
The thermal average $\langle \cdots \rangle$ is given by
\be
\langle \cdots  \rangle&=&\frac{\Tr\qty( \cdots  e^{ -\beta H})}{Z} .
\ee

The magnetization on the Nishimori line (equivalently, the spin-glass order parameter) in the thermodynamic limit is defined as 
\be
m(\beta,h)&=&\lim_{|\Lambda|\to\infty}\mathbb{E}\left[\langle \sigma_i\rangle \right]=\lim_{|\Lambda|\to\infty} \mathbb{E}\left[\langle \sigma_i\rangle^2 \right]\ge0.
\ee
We note that random fields $h_i$ on the Nishimori line (\ref{NL-condition2}) break the $\mathbb{Z}_2$ symmetry without violating the condition of the Nishimori line.
Thus, following the previous study~\cite{IS}, we define the spontaneous magnetization on the Nishimori line under the infinitesimal $\mathbb{Z}_2$-symmetry breaking fields as
\be
m_s(\beta)&=&\lim_{h\to+0} m(\beta,h).
\ee
Then, the critical inverse temperature $\beta_c^{NL}$ is defined as
\be
\beta_c^{NL}&=& \inf \left\{  \beta\ge0 \mid m_s(\beta)>0 \right\}.
\ee

The quenched free energy of the mean-field approximation on the Nishimori line is given by
\be
-\beta f_{MF}^{NL}&=& \int Dy \log2\cosh\left(  \beta\sqrt{zq+h}y+\beta^2zq +\beta^2h\right)
+\frac{z\beta^2}{4}(1-q)^2- \frac{z\beta^2q^2}{2} ,
\\
q&=&\int Dy \tanh\qty(  \beta \sqrt{zq+h}y+\beta^2zq + \beta^2h ) . \label{z-sad}
\ee
The Gibbs--Bogoliubov inequality on the Nishimori line~\cite{OO} shows that the quenched free energy of the mean-field approximation on the Nishimori line is an upper bound of the true quenched free energy
\be
f_{}^{NL} \le f_{MF}^{NL}.
\ee
For $h=0$, Eq. (\ref{z-sad}) implies that the critical inverse temperature of the mean-field approximation on the Nishimori line is given by
\be
\beta_{MF}^{NL}&=&\sqrt{\frac{1}{z}} .
\ee

Then, our result can be expressed as follows:
%%%%%%%%%%%%%%%%%%%%%%%%%%%%%%%%%%%%%%%%%%%%%%%%%%%%%%%%%%%%%%%%
%%%%%%%%%%%%%%%%%%%%%%%%%%%%%%%%%%%%%%%%%%%%%%%%%%%%%%%%%%%%%%%%
\begin{theorem}
For any positive (or vanishing) $\beta$ and $h$, the magnetization on the Nishimori line satisfies the following inequality:
\be
m(\beta,h) \le   \mathbb{E}[\tanh(\beta h_i)] + 4 z \beta^2 m(\beta,h).
\ee
Thus, there is zero spontaneous magnetization, that is,
\be
m_s(\beta)=0,
\ee
in the high-temperature region $\beta < \sqrt{1/(4z)}=\beta_{MF}^{NL}/2$, and the critical inverse temperature on the Nishimori line is bounded by 
\be
\beta_c^{NL} \ge \frac{\beta_{MF}^{NL}}{2}.
\ee
\end{theorem}
%%%%%%%%%%%%%%%%%%%%%%%%%%%%%%%%%%%%%%%%%%%%%%%%%%%%%%%%%%%%%%%%
%%%%%%%%%%%%%%%%%%%%%%%%%%%%%%%%%%%%%%%%%%%%%%%%%%%%%%%%%%%%%%%%

%%%%%%%%%%%%%%%%%%%%%%%%%%%%%%%%%%%%%%%%%%%%%%%%%%%%%%%%%%%%%%%%%
\section{Proof of Theorem 1}
Our proof is closely related to the proof of the mean-field bound for the critical temperature in the ferromagnetic Ising model using the Griffiths inequality~\cite{DLP}.

Let us fix the site $i$ and introduce the interpolating parameter $0\le t\le1$ into the Gaussian distribution of the interactions connected to the site $i$ as 
\be
J_{e} \to J_{e,t} \overset{\text{iid}}{\sim} \mathcal{N}\left(\beta t, t \right)  \: (i\in e ).
\ee
This interpolation does not break the condition of the Nishimori line~\cite{Nishimori2}.
Notably, the Gaussian distribution of all other interactions remains unchanged. 
We denote the $t$-dependence of the Gaussian distribution of the interactions as $\E[\langle \sigma_i \rangle_t]$.
The correlation function at $t=1$ coincides with the original correlation function,
\be
\E[\langle \sigma_i \rangle_1]&=&\E[\langle \sigma_i \rangle],
\ee
and the spin $\sigma_i$ is decoupled from the other spins at $t=0$,
\be
\E[\langle \sigma_i \rangle_0]&=&\E[\tanh(\beta h_i)].
\ee
On the Nishimori line, the gauge transformation shows that several correlation identities~\cite{MNC} hold for any $k$ and $l$
\be
\E[\langle \sigma_k \rangle_t \langle \sigma_k\sigma_l\rangle_t]&=&\E[\langle \sigma_k \rangle_t\langle \sigma_l \rangle_t \langle \sigma_k\sigma_l\rangle_t], \label{NL-idt-1}
\\
\E[\langle \sigma_k\rangle_t \langle \sigma_k\sigma_l\rangle_t^2]&=&\E[\langle \sigma_k\rangle_t^2 \langle \sigma_k\sigma_l\rangle_t^2],\label{NL-idt-2}
\\
\E[\langle \sigma_k \rangle_t]&=&\E[\langle \sigma_k \rangle_t^2],\label{NL-idt-3}
\\
\E[\langle \sigma_k \rangle_t \langle \sigma_l\rangle_t]&=&\E[\langle \sigma_k \rangle_t\langle \sigma_l \rangle_t \langle \sigma_k\sigma_l\rangle_t]. \label{NL-idt-4}
\ee
These correlation identities play an essential role in the following analysis.

The derivative of the correlation function with respect to $t$ is expressed as follows~\cite{MNC,ACCM}:
\be
\dv{}{t}\E[\langle \sigma_i \rangle_t]
&=&\sum_{j\sim i }\beta^2 \E[ -\langle \sigma_j\rangle_t \langle \sigma_i\sigma_j\rangle_t+\langle \sigma_i\rangle_t \langle \sigma_i\sigma_j\rangle_t^2+ \langle \sigma_j \rangle_t - \langle \sigma_i\rangle_t \langle \sigma_i\sigma_j\rangle_t]
\no\\
&=&\sum_{j\sim i }\beta^2 \E[(\langle \sigma_j \rangle_t-\langle \sigma_i\rangle_t \langle \sigma_i\sigma_j\rangle_t)^2] \label{Griffiths-NL}
\\
&\ge&0, \label{Griffiths-NL2}
\ee
where $\sum_{j\sim i }$ runs over the site $j$ connected to the site $i$ and we used integral by parts with respect to Gaussian variables in the first equality and Eqs. (\ref{NL-idt-1}), (\ref{NL-idt-2}), and (\ref{NL-idt-3}) in the second equality.
The inequality (\ref{Griffiths-NL2}) implies that the correlation function $\E[\langle \sigma_i \rangle_t]$ increases monotonically with respect to $t$, which is called the Griffiths inequality on the Nishimori line~\cite{MNC}.
By the same calculation, it is possible to prove that $\E[\langle \sigma_k \rangle_t]$ increases monotonically with respect to $t$ for any $k$.

Next, we rewrite Eq. (\ref{Griffiths-NL}) as
\be
\dv{}{t}\E[\langle \sigma_i \rangle_t]
&=&\sum_{j\sim i }\beta^2 \E[\langle \sigma_j \rangle_t^2 + \langle \sigma_i\rangle_t^2 \langle \sigma_i\sigma_j\rangle_t^2-2\langle \sigma_i\rangle_t \langle \sigma_j \rangle_t ],
\ee
where we used Eq. (\ref{NL-idt-4}).
Using a trivial inequality, $-2\E[\langle \sigma_i \rangle_t \langle \sigma_j\rangle_t] \le\E[ \langle \sigma_i \rangle_t^2+\langle \sigma_j\rangle_t^2]$, we obtain 
\be
\dv{}{t}\E[\langle \sigma_i \rangle_t]
&\le&\sum_{j\sim i }\beta^2 \E[\langle \sigma_j \rangle_t^2 + \langle \sigma_i\rangle_t^2 \langle \sigma_i\sigma_j\rangle_t^2 +\langle \sigma_i \rangle_t^2+\langle \sigma_j\rangle_t^2 ]
\no\\
&\le&\sum_{j\sim i }2 \beta^2 \E[\langle \sigma_i\rangle_t^2 + \langle \sigma_j \rangle_t^2    ]
\no\\
&=&\sum_{j\sim i }2 \beta^2 \E[\langle \sigma_i\rangle_t + \langle \sigma_j \rangle_t    ],
\ee
where we used Eq. (\ref{NL-idt-3}) in the equality.
Then, the Griffiths inequality on the Nishimori line (\ref{Griffiths-NL2}) implies
\be
 \E[\langle \sigma_i\rangle_t ]\le  \E[\langle \sigma_i\rangle_1 ],
\ee
which leads to 
\be
\dv{}{t}\E[\langle \sigma_i \rangle_t]
&\le&\sum_{j\sim i }2 \beta^2 \E[\langle \sigma_i\rangle_1 + \langle \sigma_j \rangle_1 ].
\ee
Thus, by integrating $t$ from $0$ to $1$ and using $\E[ \langle \sigma_i \rangle_1]=\E[ \langle \sigma_i \rangle]$, we have
\be
\E[\langle \sigma_i \rangle]&\le&\E[\langle \sigma_i \rangle_0] +\sum_{j\sim i }2 \beta^2 \E[\langle \sigma_i\rangle + \langle \sigma_j \rangle ]
\no\\
&=&\E[\tanh(\beta h_i)] +\sum_{j\sim i }2 \beta^2 \E[\langle \sigma_i\rangle + \langle \sigma_j \rangle ]. \label{main-ineq}
\ee

Finally, we take the thermodynamic limit in Eq. (\ref{main-ineq}).
The Griffiths inequality on the Nishimori line guarantees the existence of the thermodynamic limit for the correlation function, $\lim_{|\Lambda|\to\infty}\E[\langle \sigma_i \rangle]$, under both free and fixed boundary conditions on the Nishimori line~\cite{MNC}.
Furthermore, for these boundary conditions, the Griffiths inequality on the Nishimori line enables us to prove that the correlation function exhibits translational invariance in the thermodynamic limit, $\lim_{|\Lambda|\to\infty}\E[\langle \sigma_i \rangle]=\lim_{|\Lambda|\to\infty}\E[\langle \sigma_j \rangle]$, similar to ferromagnetic systems~\cite{FV}.
Thus, by $\sum_{j\sim i }=z$, we obtain
\be
m(\beta,h)&\le& \E[\tanh(\beta h_i)] +\sum_{j\sim i }4 \beta^2 m(\beta,h)
\no\\
&=& \E[\tanh(\beta h_i)] +4 z \beta^2 m(\beta,h),
\ee
in the thermodynamic limit. This completes the proof of Theorem 1.

%%%%%%%%%%%%%%%%%%%%%%%%%%%%%%%%%%%%%%%%%%%%%%%%%%%%%%%%%%%%%%%%%%
%%%%%%%%%%%%%%%%%%%%%%%%%%%%%%%%%%%%%%%%%%%%%%%%%%%%%%%%%%%%%%%%%%
\section{Discussions}
We proved that the true critical inverse temperature on the Nishimori line is bounded by $\beta_c^{NL} \ge \beta_{MF}^{NL}/2$.
Our proof is based on the proof of the mean-field bound for the critical temperature in the ferromagnetic Ising model~\cite{DLP}, where the Griffiths inequality~\cite{Griffiths3,KS,Ginibre} play an important role.
Correspondingly, the key component of our proof is the Griffiths inequality on the Nishimori line~\cite{MNC}, which enables us to address the randomness in spin glass models. 

Our bound is a weak correspondence of the mean-field bound for the critical temperature in ferromagnetic spin models.
It remains an important future problem to investigate whether the corresponding mean-field bound on the Nishimori line, $\beta_c^{NL} \ge \beta_{MF}^{NL}$, holds.

In this study, we only considered Ising spin glass models with Gaussian randomness on the Nishimori line.
It would be interesting to investigate whether a corresponding mean-field bound for the critical temperature holds for other types of randomness, such as the $\pm J$ model, on the Nishimori line.

In ferromagnetic spin models, there exists a mean-field bound for the magnetization~\cite{Thompson,Pearce,Slawny,Newman,TH}, which is stronger than the mean-field bound for the critical temperature.
An interesting problem to be addressed in the future is to obtain a corresponding mean-field bound for the magnetization on the Nishimori line.

%%%%%%%%%%%%%%%%%%%%%%%%%%%%%%%%%%%%%%%%%%%%%%%%%%%%%%%%%%%%%%%%%%
%\begin{acknowledgment}
The authors are grateful to Hal Tasaki for useful comments.
This study was supported by JSPS KAKENHI Grant Nos. 24K16973 and 23H01432.
Our study received financial support from the Public\verb|\|Private R\&D Investment Strategic Expansion PrograM (PRISM) and programs for bridging the gap between R\&D and IDeal society (Society 5.0) and Generating Economic and social value (BRIDGE) from the Cabinet Office.
%\end{acknowledgment}

%%%%%%%%%%%%%%%%%%%%%%%%%%%%%%%%%%%%%%%%%%%%%%%%%%%%%%%%%%%%%
%%%%%%%%%%%%%%%%%%%%%%%%%%%%%%%%%%%%%%%%%%%%%%%%%%%%%%%%%%%%%%%%%%%%%%%%%%%%

\end{document}